# Defect dependent dynamic nanoindentation hardness of copper up to 25 000 s$^{-1}$

Hendrik Holz[a], Lalith Kumar Bhaskar[a], Tobias Brink[a], Dipali Sonowane[a], Gerhard Dehm[a], James P. Best[a], Rajaprakash Ramachandramoorthy[a]

[a]Max Planck Institute for Sustainable Materials, Max-Planck-Straße 1, 40237 Düsseldorf, Germany

## Abstract

Metals exhibit an upturn in strength at strain rates of approximately 1 000 s$^{-1}$ – 3 000 s$^{-1}$, governed by rapid dislocation multiplication, interactions and storage. This phenomenon is strongly influenced by the initial dislocation density before testing. However, the role of immobile dislocations arranged in low-angle grain boundaries (LAGBs) on deformation under such extreme conditions remains unexplored, despite their ubiquity in engineering materials. Here, we employ high strain rate nanoindentation targeted at an LAGB with tilt and twist components in copper crystals with different dislocation densities. We demonstrate that Taylor hardening remains valid over a wide range of strain rates. It was found that the influence of LAGBs on mechanical properties is within the scatter of the measurements. However, slip traces of indents close to the LAGB suggest that the LAGB acts as a barrier to dislocations. Molecular dynamics simulations further confirm these findings. The measured activation volume and low strain rate re-indentation onto indents performed at different higher strain rates give insights into the deformation mechanism. This work provides new insight into the interplay between microstructure and high strain rate deformation.

## Graphical Abstract

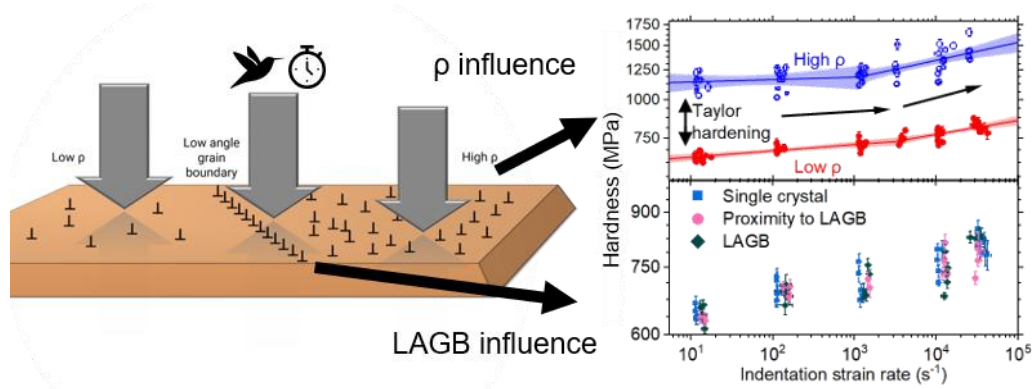

## Highlights

- High strain rate nanoindentation applied to copper single and bi-crystal with LAGB
- Strength upturn observed above 3 000 s$^{-1}$ in both high and low dislocation densities
- LAGB acts as a dislocation barrier but seems not to affect the upturn strain rate
- Activation volume decreases in the upturn regime
- Reloading indents indicate different microstructure evolution for low and high strain rate tests

# Introduction

Understanding the mechanical behavior of materials during dynamic deformation is important for applications involving high strain rate loading. One example is machining, where the large strain and strain rates significantly alter the mechanical behavior and microstructure of the material [1,2]. Therefore, there is sustained interest in testing several material types at high strain rates, historically mostly at the macroscale [3,4] and since the early 2000s also at the micron scale, pioneered by Beake et al. [5], with recently growing effort [6,7]. Metals exhibit higher strengths with increasing strain rate [8]. An upturn in strength at strain rates above 1 000 $s^{-1}$ has been consistently observed, where the strength increase becomes significantly larger [3,9].

Typically, the upturn in strength was attributed to dislocation-phonon drag [10], but recently, there is evidence that the contribution of phonon drag could only be significant at strain rates above $10^5$ $s^{-1}$, when the majority of dislocations can move close to the material's wave speed [11]. At more moderate high strain rates, Follansbee et al. developed a physically based model using the temperature, strain rate, and a structure dependent mechanical threshold strength as state variables [9,12]. The model suggests that dislocations nucleate rapidly in metals at high strain rates, while dynamic recovery becomes suppressed [8,9,13]. This in turn leads to a "rate sensitivity of structure evolution" [9]. Gray also came to the same conclusion from post mortem dislocation morphology analysis [3]. Direct proofs for these theories are still missing because currently existing atomic resolution microscopes cannot resolve the dislocation motions at these speeds [14].

In recent times, with advances in nanoindentation instrumentation at high strain rates, similar results were observed in micro-scale experiments as well. For example, high strain rate nanoindentation in aluminum revealed differences in subgrain formation and formation of microbands compared to quasi-static tests [15,16]. Also, a detailed post-deformation analysis of high strain rate indentation on molybdenum revealed an increase in dislocation density beyond the hardness upturn in line with was observed from macroscale experiments [17]. These results highlight the coupled roles of strain rate and dislocation density evolution in governing the mechanical response. In line with this, Fan et al. used discrete dislocation dynamics (DDD) and molecular dynamics (MD) simulations to investigate the relationship between initial dislocation density and the strain rate at which the upturn in strength occurs [18]. An analytical model was derived, describing the forest hardening as well as the strain rate hardening regime using, among others, strain rate and dislocation density as parameters. Various experimental datasets collapse onto this model. The upturn in yield is then dependent on the strain rate and the dislocation density in the theory.

An open question is how local dislocation structures, such as those forming low-angle grain boundaries (LAGBs), influence this behavior. Since LAGBs occur in most engineering materials, including in nominal single crystals as subgrain boundaries [19], it is essential to investigate their influence on high strain rate deformation. So far, macroscale mechanical testing does not allow the testing of isolated microstructural features, but only whole samples at once, containing a variety of defects. Here, micromechanics has an advantage since experiments can be performed on specific isolated defects with a high spatial resolution. A test can thus be targeted at a selected microstructure component, for example a grain boundary. Many nanoindentation experiments in proximity to high angle grain boundaries (HAGB) were performed mostly in body centered cubic metals [20]. Jakob et al. found that an increase in hardness in the proximity of a HAGB depends on the orientation of the pyramidal indenter tip and is only found if the orientation is such that the deformation patterns along the crystallographic orientations are facing towards the HAGB [21]. For face centered cubic (fcc) metals, data in the literature is sparser. In copper, it was found that the hardness in proximity to HAGBs can result in an increase of ~1.5 times the bulk hardness [22]. The increased hardness was found up to a distance of 2 µm from the HAGBs, even though the indentation depth was only around 350 nm [22]. While there is a significant amount of work focused on HAGBs, comparatively few studies have examined LAGBs. Some of the micromechanical experiments on LAGBs use either nanoindentation and post deformation high resolution electron backscatter diffraction (EBSD) [23], *in situ* experiments in the transmission electron microscope (TEM) [24], or compression experiments on micropillars that contain a LAGB [25,26]. They observed that a LAGB can interact with dislocations by transmission, absorption and pile-up formation. This in turn leads to the strengthening of the materials. But so far, all these tests were performed at quasi-static strain rates, and no such tests have been conducted in the high strain rate range. Extending such investigations to the high strain rate regime remains largely unexplored, primarily due to the limited availability of nanoindentation systems capable of operating at high strain rates. Due to the capability of LAGBs to act as dislocation sources and dislocation sinks under mechanical stress [23], these mechanisms could shift the critical strain rate for the upturn in hardness in either direction or increase the strain rate sensitivity similar to observations on samples with a high amount of HAGBs in copper [26]. Understanding the response of LAGBs under these conditions is therefore of particular interest, as it may reveal new insights into their role in dislocation interactions and rate-dependent deformation mechanisms.

Therefore, in this work, we look into the influence of dislocation density and of a LAGB on the mechanical properties at constant strain rates from 10 $s^{-1}$ - 25 000 $s^{-1}$ in a copper single crystal by nanoindentation. Using a custom modified piezo-based system with advanced signal processing [17], we test whether LAGBs and different dislocation densities affect hardness or the critical strain rate for the strength upturn, and support our findings with MD simulations.

## Methods

### Experimental

The bulk sample was a bicrystal grown using the Bridgman process. Two slices of 2 mm thickness were cut by electro-discharge machining. The samples were mechanically polished using 600 grid SiC paper to create a flat and parallel surface, followed by grinding steps with 800, 1200, 2400, and 4000 grids. One slice was polished with 3 µm polishing solution, followed by electro polishing (from now on called "high dislocation density sample"). The other slice was polished with 3 µm, then 1 µm polishing solution and an oxide polishing suspension (OPS), followed by electropolishing (from now on called "low dislocation density sample"). The electropolishing was done for 5 s at 8 V using the electrolyte D2 by Struers Inc.

EBSD measurements were performed on an Auriga cross beam focused ion beam (FIB) - scanning electron microscope (SEM) (Zeiss AG) using a Hikari EBSD camera (EDAX) with an acceleration voltage of 30 keV and an aperture size of 120 µm. A FIB cross section across the LAGB confirmed that the LAGB runs perpendicular to the surface. Using this information and the data gathered from the EBSD, the twist and tilt components of the LAGB were calculated using the equations from Ref. [27]. From the EBSD data the geometrically necessary dislocation (GND) density for both samples was determined using OIM analysis version 8.6.0101 based on Ref. [28]. GNDs are all dislocations required to accommodate the lattice curvature caused by plastic deformation. Unlike statistically stored dislocations, which accumulate randomly, GNDs arise from orientation gradients that EBSD can directly measure.

Nanoindentation experiments were performed using an *in situ* nanoindenter (Alemnis AG) that was customized with a piezotube actuator and a piezoelectric load cell in order to be able to perform high constant strain rate nanoindentation as described in Ref. [17]. For this, the signals of the load cell were amplified using a high-impedance charge amplifier (Alemnis AG). The voltage applied to move the piezo actuator was supplied by a high-speed and high-voltage amplifier (WMA-200 Falco Systems). The resulting displacement is captured by strain gauges that are attached to the piezo actuator. The measured data requires significant postprocessing. First, the load and displacement data have a time shift of ~5 µs, owing to the different electronics used to capture them, which need to be synchronized. Furthermore, the load sensor has a time constant $\tau$ associated with it, which indicates the speed at which the system reacts to dynamic changes. The time constant was characterized in Ref. [17]. Assuming a first order time constant, the corrected load is $P_{\text{corr}} = P_{\text{meas}} + \dot{P}\tau$ with $P_{\text{meas}}$ being the measured load and $\dot{P}$ the loading rate [29]. Finally, the machine dynamics, including inertia forces and dampening of the system, need to be accounted for. This was done by an iterative quadratic-fit method that uses lower-rate data to generate mock loads and correct higher-rate responses for load-cell inertia and damping. For detailed information, the reader is referred to Ref. [17].

The indentation was performed on 5 different positions on both samples. In the bulk of both grains, at the grain boundaries, and in proximity to the LAGB at roughly 8 µm distance in both grains. The total indentation depth was 1.2 µm, and the lateral spacing of the indents was at least 20 times the indentation depth to avoid interactions of the plastic zones underneath the indents [30]. In order to analyze the hardness, the iterative method by Merle et al. based on the loading portion of the load-displacement curve was used [31], since the unloading data is influenced by resonance frequencies at strain rates above 1 000 s$^{-1}$. A reduced modulus of 136.5 GPa was measured at low strain rates and used for the calculation of the hardness using the iterative method. To avoid the influence of the indentation size effect, the hardness data were averaged from the same depth range of 600 nm – 800 nm for all strain rates. It should be noted that using laser interferometers positioned laterally to the piezotube actuator, a ~10% lateral movement during unconstrained in-air tests was identified. It would be further diminished during indentation due to the sample constraint. Expectedly, the lateral forces calculated from the piezoelectric load cell data across all the tests reported here ≤ 6.5% of the maximum loads. With increasing strain rate, the absolute and relative values of lateral loads compared to the total normal indentation load decrease, as shown in Table S1 in the supplementary material; therefore, it is clear that they are not contributing to the hardness upturn.

To probe microstructural changes induced by high strain rate loading, two-step reloading experiments were performed. In the first step, indents were made at strain rates of 10 $s^{-1}$ – 25 000 $s^{-1}$. Immediately afterward, a second indent was applied at 10 $s^{-1}$ at the same location. The reloading indents were analyzed by the Oliver-Pharr method [32] to obtain hardness.

## Atomistic simulations

We used some supplemental molecular statics and MD simulations to elucidate the dislocation structure and behavior of the LAGB. The simulations were performed with LAMMPS [33] using an embedded atom method (EAM) potential for copper [34]. First, we constructed a cylindrical bicrystal with the same macroscopic grain boundary parameters as found in the experiment to reproduce the studied LAGB. This sample had a diameter of 30 nm and a length of 30 nm, the LAGB being in the center with its normal aligned along the cylinder's axis. The structure was minimized to obtain the dislocation network of the LAGB. The dislocations were analyzed with the dislocation extraction algorithm (DXA) [35,36] implemented in Ovito [37].

We studied the dislocation transmission by inserting a full dislocation into one crystallite using Atomsk [38] and minimizing the sample. Expectedly, the full dislocation splits into Shockley partials. We then ran a quasi-athermal MD simulation with a time integration step of 2 fs and a Nosé-Hoover thermostat at 50 K. We fixed the top and bottom boundaries over a length of three times the EAM potential's cutoff range. The boundaries were continuously moved closer together to achieve a compression with an engineering strain rate of $10^8$ $s^{-1}$. As a result, the dislocation was driven towards the LAGB.

## Results & Discussion

## Microstructure analysis

The microstructures of the high and low dislocation density samples are shown in Figure 1. The inverse pole figure map reveals two different sub-grains with similar crystallographic orientations for both samples. The surface normal of one sub-grain is oriented close to the [6 4 11] crystal direction and the surface normal of the other sub-grain is close to [7 5 9]. The high dislocation density sample contains a GND density of $(5.3 \times 10^{13} \pm 1.9 \times 10^{13})$ m$^{-2}$ while the low dislocation density sample has $(1.9 \times 10^{13} \pm 0.8 \times 10^{13})$ m$^{-2}$. The boundary between the two grains exhibited a misorientation of 12°, which is the smallest rotation to make one grain coincide with the other one, while considering all crystallographic symmetries. This misorientation consists of an 8° twist component and a 9.6° tilt component.

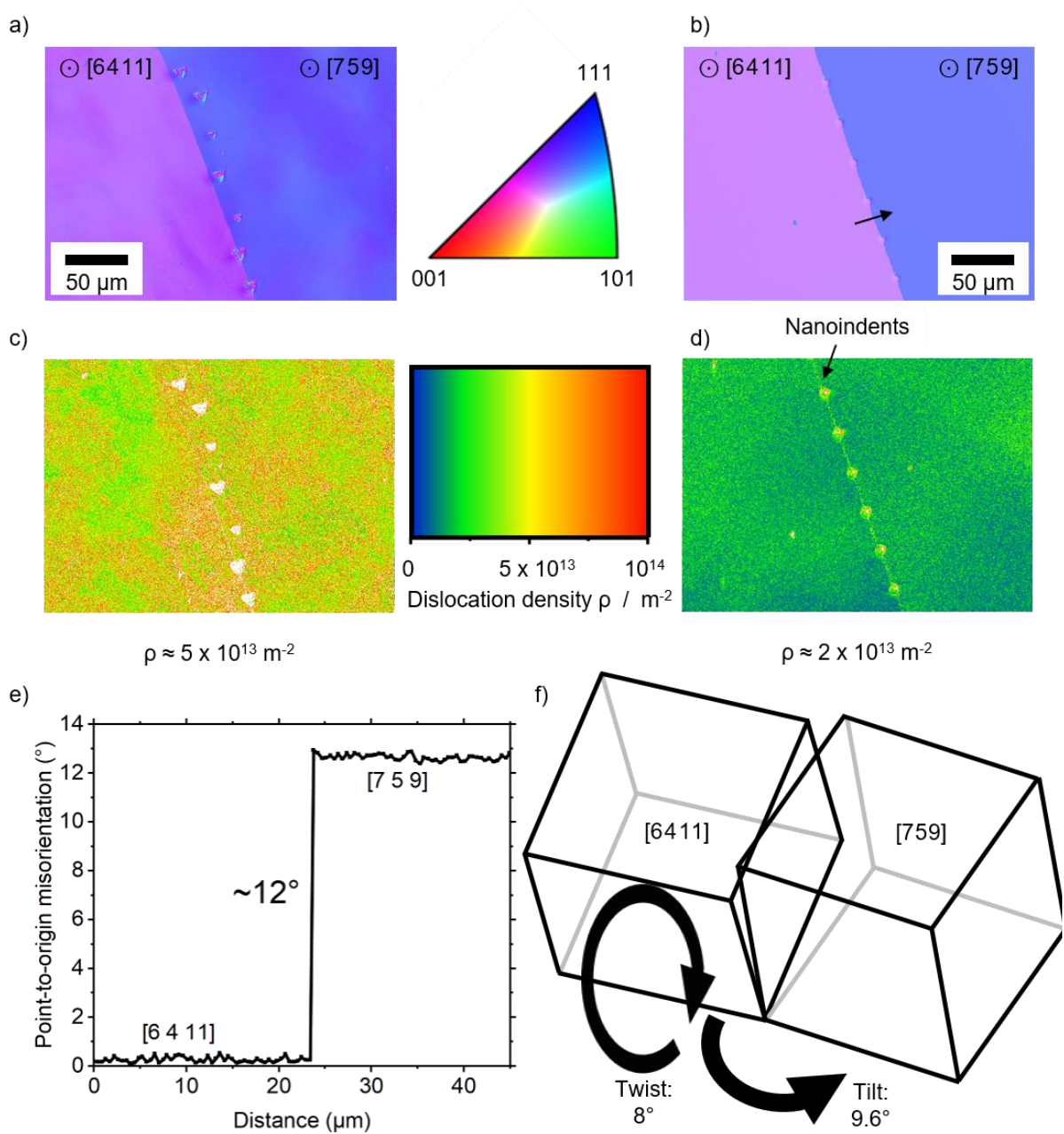

*Figure 1: Microstructure of the investigated copper samples. Orientation maps of the (a) high and (b) low dislocation density samples. (c,d) Maps of geometrically necessary dislocation density. (e) Misorientation profile across the low-angle grain boundary (LAGB) marked in (b). (f) Schematic representation of the tilt and twist components of the LAGB.*

To further investigate the boundary structure, molecular statics simulations were performed. These revealed that the grain boundary indeed consists of a network of dislocations and is therefore a LAGB. The dislocations are predominantly full dislocations and Shockley partials with both edge and screw character, as illustrated in Figure 2. These results are expected for mixed LAGBs [39]. The combination of EBSD and MD thus confirms that the investigated boundary is a structurally complex LAGB, providing a suitable model system for assessing the influence of immobile dislocations on the mechanical properties at high strain rates.

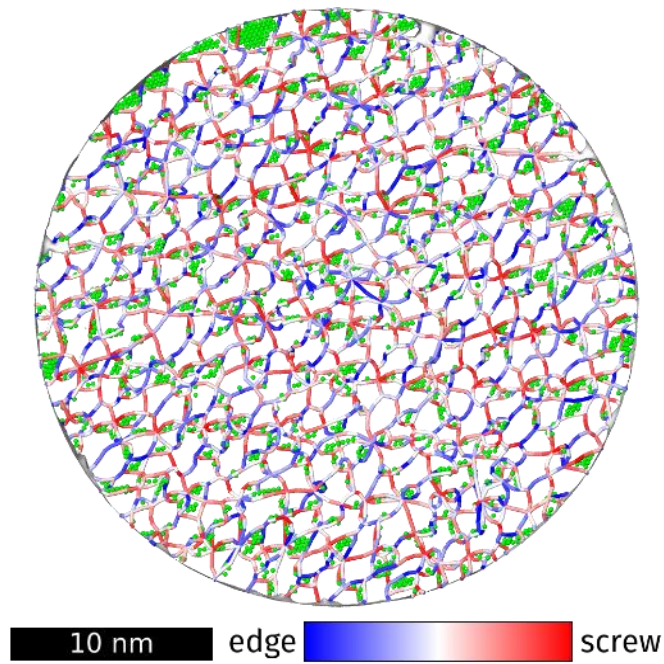

*Figure 2: Molecular statics simulation of the LAGB's dislocation structure. The image shows a view from the direction normal to the grain boundary. Lines are dislocations as identified by DXA, with the color coding indicating the degree of edge or screw character. The green atoms were identified as belonging to stacking faults. All other atoms are not shown. From DXA we obtain a total dislocation length of 1.8 µm. With a radius of 15 nm of the circular grain boundary, we obtain an area density of 2.57 x $10^9$ $m^{-1}$ for the dislocations.*

## Mechanical response

### Indentation data

The nanoindentation data (Figure 3) confirm that constant indentation strain rates were maintained up to depths of 800 nm for all targeted strain rates, up to the highest rate of ca. 25 000 s$^{-1}$ conducted in this work. The initial drop in strain rate in the first few hundred nanometers is due to the indenter not being in contact at the start of the experiment. The loading portions of the load-displacement curves of the single crystal indentation, shown in Figure 3 (b), exhibit the expected increase in load with increasing strain rate, characteristic of strain-rate hardening. The small standard deviations highlight the high reproducibility of the experiments. This is also the case for the load displacement curves of the indents conducted on the LAGB and close to it (Figure S1).

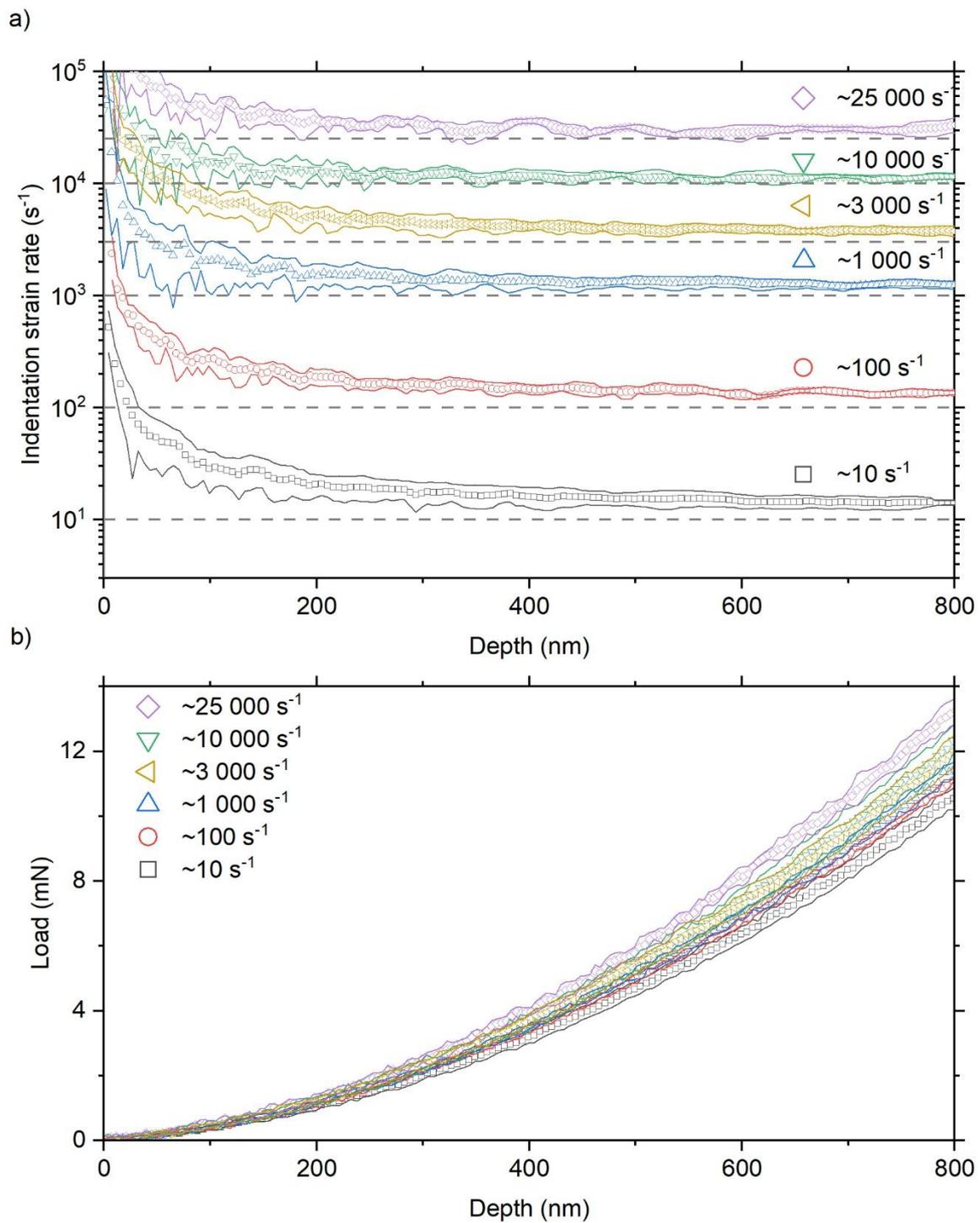

*Figure 3: Representative data of nanoindentation experiments in a single grain after postprocessing corrections as described in the methods section. The circles represent the average of at least 3 experiments per strain rate and the lines show the standard deviation. (a) Indentation strain rate over depth. (b) Load-depth curve.*

Hardness values were extracted from the loading portion of the curves using the Merle-Higgins-Pharr method [31], averaged between 600 nm - 800 nm depth to minimize indentation size effects. At that point, the estimated plastic zone size should be 9 µm – 12 µm in radius, assuming the plastic zone size is 15 times the indentation depth [30]. The hardness from the high and low dislocation density samples at specific locations in a single grain, 8 ± 0.9 µm away from the LAGB and at the LAGB, are plotted as a function of strain rate in Figure 4 (a). In Figure S2 in the supplementary materials, a comparison of the data with the literature is shown.

**Influence of the LAGB**

Surprisingly, the location of the indents in relation to the LAGB does not influence the measured hardness values within the range of scattering of the data. In Figure S3, the hardness as a function of strain rate at different positions is plotted. Therefore, no change in the strain rate at which the upturn in hardness occurs is measured. Nevertheless, SEM micrographs reveal an interaction of the dislocations with the LAGB. For indents in proximity to the LAGB, the slip lines stop at the LAGB as visible in Figure S4 (a), whilst indents on the LAGB show slip lines along the primary (111)⟨110⟩ slip systems in the respective single grains as seen in Figure S4 (b) in the supplementary materials. Only these slip systems were found to be activated in the single grains and were analyzed using the method from Ref. [40]. It is worth noting that the slip lines are facing towards the LAGB, which was found to be a requirement in HAGBs to observe a change in the mechanical response [21]. Thus, the LAGB acts as an obstacle to dislocation motion, but its influence is too small to measurably affect hardness under our indentation conditions. This is particularly surprising since other research revealed strengthening from the introduction of LAGBs, for example, in micropillars [25], even though the observation was based on just one specimen. In a medium entropy alloy with fcc structure, an influence of a LAGB on the mechanical behavior was found [23], although the hardness was not reported, a lack of pop-ins for indents performed close to the LAGB confirmed the behavior [23]. A similar finding for fcc metals using DDD simulations was made, where an increase in dislocation density of the LAGB led to an increase in transmission resistance [41]. A more sensitive experimental technique, such as micropillar compression containing the grain boundary, could provide an alternate and possibly slightly better insight into the mechanical influence of the probed LAGB, similar to what has been done with high angle grain boundaries [42]. However, in our case, the out of plane orientations of the crystals vary, leading to different FIB milling rates and therefore making the manufacturing of micropillars cumbersome.

We instead analyzed the atomic-scale mechanisms of the LAGB using MD simulations. As depicted in Figure 2, the dislocation spacing is on the nanometer scale. Using a similar derivation as for Taylor hardening (see Supplemental Note 1), we predict a maximal additional shear strength of 6.3 GPa for transmission of the dislocation through the GB. Therefore, the dislocation transmission across the LAGB is not achieved due to this high additional shear strength requirement. For a more detailed analysis using the MD simulation, we inserted a full dislocation into the bulk of one grain of the bicrystal model, which was shown in Figure 2. Then, we compressed the cylinder in the direction normal to the LAGB (Figure 5). The hardness was calculated from the uniaxial stress assuming a constraint factor of 3 and the uniaxial stress from resolved shear stress $\tau_{RSS}$ using a Schmid factor of ~0.5. Immediately, the dislocation splits into two Shockley partials and we observe some initial pinning of the dislocation line. The depinning stress is quite high, but significantly below the transmission stress. We therefore exclude that any artificial dislocation pinning influences the transmission. As expected from a Taylor-like hardening model, we only observe bowout of partials in the experimental stress ranges (Figure 5(b)). At shear stresses around 1 GPa, transmission events take place. This is below the estimation using a Taylor-like model due to dislocation reactions (Figure 5(d),) which was shown for dislocation LAGB interaction using DDD in fcc as well as in bcc metals [41,43,44]. The required stresses nevertheless far exceed the experimental hardness of ≤ 1.5 GPa. This correlates with the missing slip traces across the

LAGB in the experiments (Figure S4(a)) and also to results in the literature of DDD simulations where the LAGB was a strong obstacle to dislocations [41].

The results for copper HAGBs in Ref. [22] show grain-boundary-related hardening exceeding the scatter of our present experimental data. Surprisingly, we do not observe any such effects near the LAGB, despite the lack of dislocation transmission across it. In Ref. [22], several hypotheses were proposed for the hardening. The elastic anisotropy between the abutting crystallites at the HAGB might lead to attraction or repulsion of the dislocations, affecting the hardness. Additionally, the slip transfer mechanisms might also be anisotropic. In the case of LAGBs the elastic anisotropy is much lower than for HAGBs and we observe no transmission in either direction. Hardening due to anisotropy can therefore mostly be excluded in our case. Additionally, grain boundaries might also act as sinks for dislocations or sweep over large regions of the crystal during annealing in some samples, thereby depleting the vicinity of dislocations and affecting the hardness. Here, we measured the GND density and found no such dislocation density gradient between the bulk and the grain boundary.

### Influence of the bulk dislocations

The high-dislocation-density sample shows invariably higher hardness than the low-density sample, consistent with Taylor hardening. The increase agrees quantitatively with the Taylor hardening factor [45] estimated from GND densities using $\sqrt{\rho_{\text{high}}/\rho_{\text{low}}}$, where $\rho_{\text{low}}$ and $\rho_{\text{high}}$ correspond to the low and high dislocation density, respectively, as measured by EBSD, as can be seen in Figure S5. Furthermore, the travel distance of visible slip lines from indents is higher for the low dislocation density sample (Figure S6) compared to the high dislocation density sample, even though the indentation depth is kept constant. This indicates that there are fewer interactions between the dislocations and thus they can move further into the material, resulting in the reduction of hardness.

The strain-rate dependence of hardness reveals two distinct regimes visible in Figure 4 (a). A two-segment exponential piecewise fit with a continuity condition at the breakpoint and the position of the breakpoint as a variable was used to exclude subjectivity while finding the critical strain rate where the upturn happens. The fit is further explained in Supplemental Note 2. This exponential fit corresponds to linear fits in the log-log plot. As a comparison, we also fitted a single strain rate dependency over the whole range, intentionally ignoring the upturn. A comparison between the root mean square deviation (RMSD) values for a single fit over the whole strain rate range and the piecewise fits with two regimes is shown in the Supplemental Table 2: The two-regime function offers a better fit to the data. At strain rates up to ~3 000 s$^{-1}$, the strain-rate sensitivity is low with 0.008 ± 0.005 and 0.020 ± 0.003 for the high and low dislocation density samples, respectively. Starting from ~3 000 s$^{-1}$, both samples exhibit a clear upturn in hardness, with slopes increasing to 0.06 ± 0.02 and 0.041 ± 0.008 for the high and low dislocation density samples, respectively. According to Fan et al., there should be a shift of the upturn to higher strain rates with higher dislocation density [18], which is not observed here. Presumably, the reason for this is the comparatively small difference in dislocation densities, the gap between the strain rates in the upturn regime, and an insufficient number of experiments to quantify a shift of the critical strain rate at which the upturn occurs. Additionally, even in the simulations of Fan et al. the scatter of critical dislocation densities for a given critical strain rate is in the range of two orders of magnitude [18] as visible in Figure 4 (b). A more detailed analysis would require samples with vastly different dislocation densities which is out of scope for this publication and will be explored in the future. Nonetheless, a comparison of the dislocation density and the strain rate at which the upturn starts shows that our data lies within the 95 % prediction band of the modeled data from Fan et al. (Figure 4 (b)). The error bars in strain rate of our data were chosen to be the resolution of the data we have within the upturn regime. Importantly, indents at or near the LAGB do not alter the critical strain rate of the upturn in an observable range, indicating that mostly mobile dislocations control the transition. This implies that a single LAGB has no significant strengthening effect for materials under loading. However, it is worth noting that the high scatter of the data prohibits a precise identification of the strengthening effect.

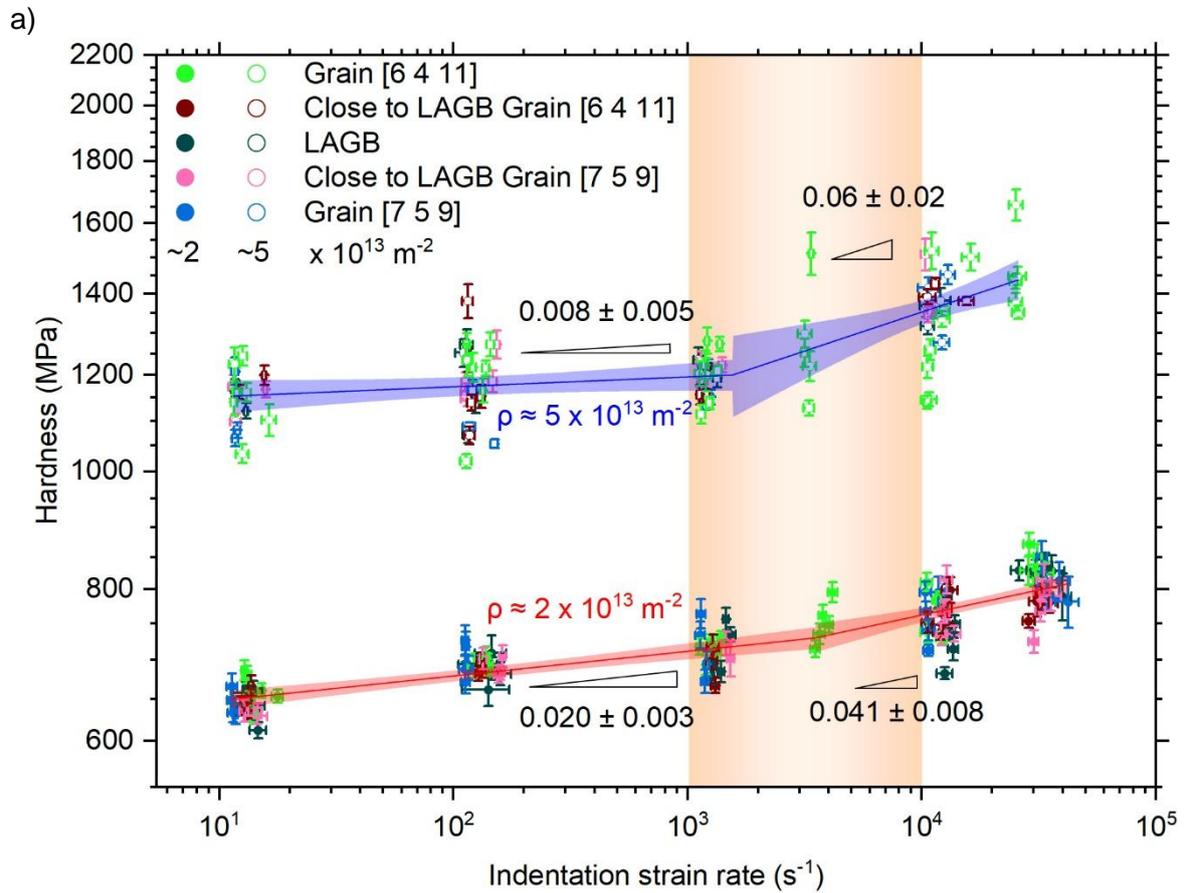

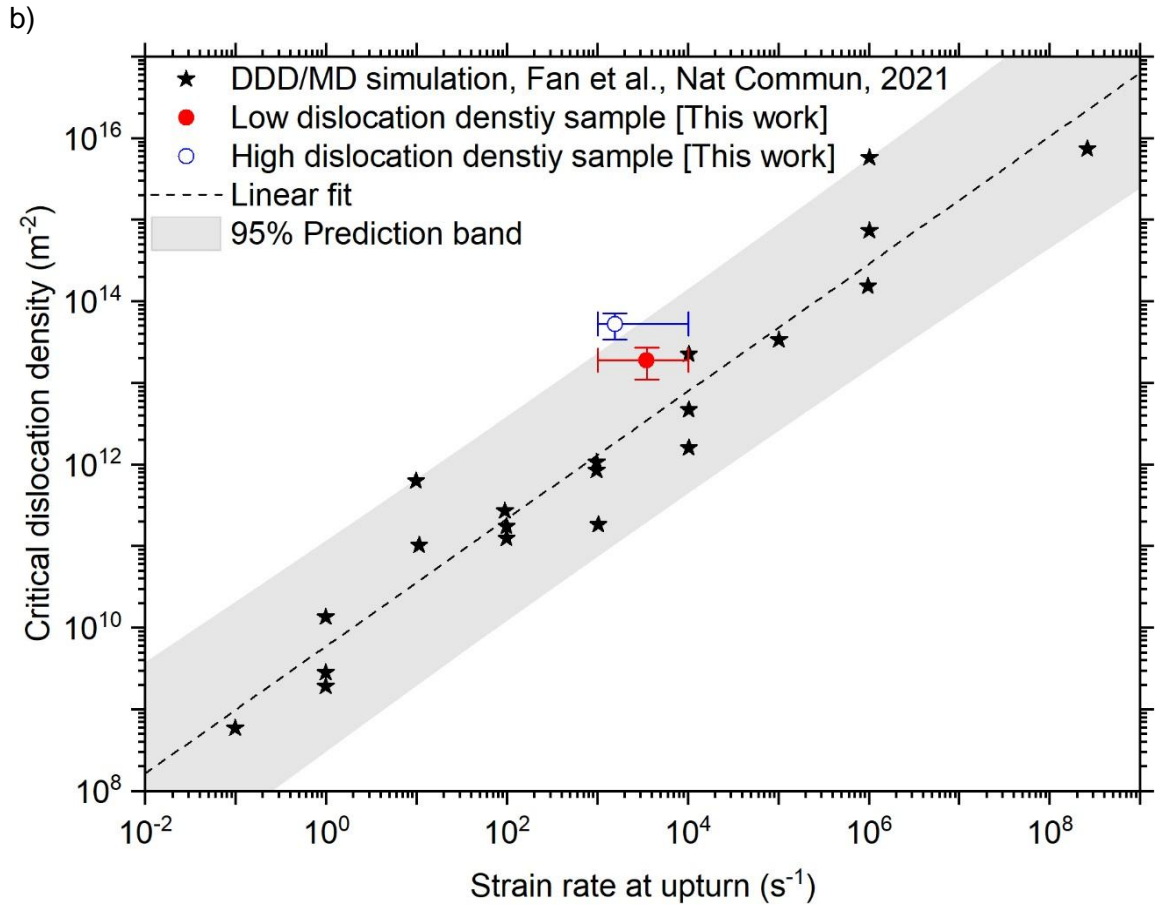

Figure 4: Rate sensitivity of copper with varying dislocation density. (a) Inverse Norton plots of nanoindentation tests performed in single grains (10 µm from the LAGB) and at the LAGB, for high (open circles) and low (filled circles) dislocation density samples. Piecewise exponential fits with 95% confidence bands are shown; the orange bar highlights the strain-rate range of the hardness upturn. (b) Comparison of experimentally measured upturn strain rates as a function of critical dislocation density with predictions from Fan et al. [18].

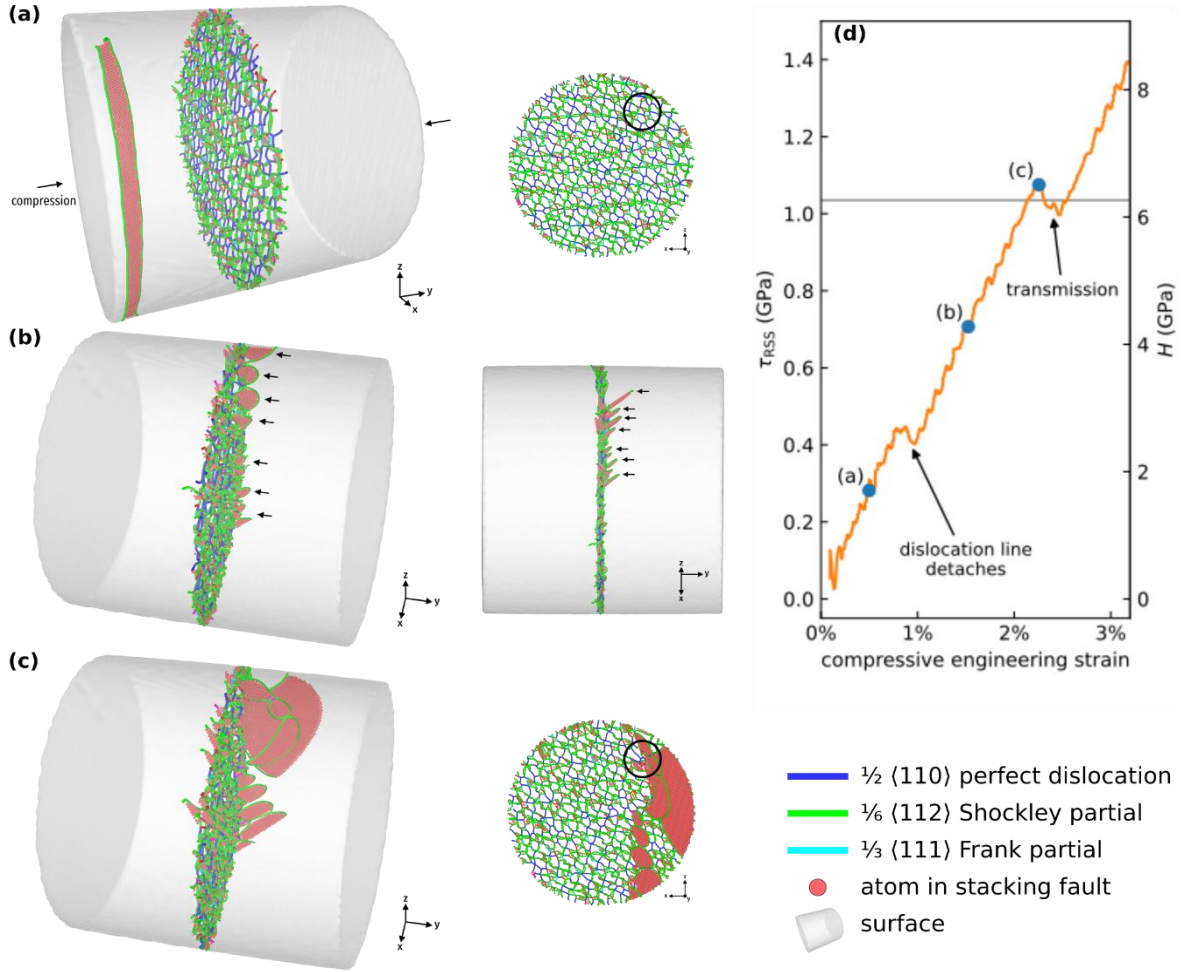

Figure 5: *Dislocation–LAGB interaction studied by MD simulation. A pre-inserted pair of partial dislocations is driven towards the LAGB by uniaxial compression. (a) The inserted partial dislocation pair is initially pinned near the surface/boundary, likely due to the small sample geometry. Only dislocation lines as identified by DXA and atoms that are part of stacking faults are shown. (b) After detachment, the dislocations then get blocked by the LAGB. As expected, we observe bow-outs in between the LAGB dislocations (marked by arrows). Note that the bow-outs are not located on the same plane due to twist components of the LAGB. (c) At sufficiently high stresses, leading partials are transmitted. The marked regions in the frontal views of the LAGB in (a) and (c) show that the LAGB's dislocation networks have changed during transmission, indicating a dislocation reaction. (d) The corresponding stress–strain graph. The hardness on the second ordinate was estimated from the uniaxial stress using a constraint factor of three.*

From the hardness, and using the strain rate sensitivity *m*, the apparent activation volume *V* can be calculated using:

$$V = \sqrt{3}\frac{kT}{m\sigma_f} \cong 3\sqrt{3}\frac{kT}{mH}, \tag{1}$$

where *k* is the Boltzmann constant, *T* the temperature and $\sigma_f$ the flow stress. The latter can be related to the hardness *H* using a constraint factor that is usually assumed to be 3 for metals. Using the Burgers vector *b* = 2.556 Å for a full dislocation in copper [46], the activation volume can be calculated and related to deformation processes. The resulting activation volumes for the high and low dislocation density samples are shown in Figure 6. The data points at the strain rates in the proximity of 3 000 s$^{-1}$ are displayed twice: calculated once using the strain rate sensitivity of the lower strain rate regime and calculated another time using the higher strain rate regime ones. For both regimes, the activation volumes are stable. In the pre-upturn regime the samples show an activation volume of (93± 5) b$^3$ and (136± 9) b$^3$ for the low and high dislocation density sample respectively which is the lower bound for dislocation-dislocation interactions in fcc metals [47], but lower than values reported for cross-slip based deformation in copper, which happens between 300 b$^3$ - 400 b$^3$ [48–50]. This suggests that in the early regime, plasticity is not dominated by cross-slip, but rather by short-range dislocation interactions.

In the regime of the upturn, the activation volume decreases to (40 ± 2) b$^3$ or (16 ± 1) b$^3$ for the low and high dislocation density sample respectively. Such a reduction points toward the activation of mechanisms involving fewer atomic units, which is consistent with dislocation motion being increasingly confined or hindered at high strain rates. This is consistent with observations in previous high strain rate nanoindentation studies performed in aluminum and molybdenum [15–17]. The difference between the two samples indicates that dislocation density directly affects the efficiency of these mechanisms, in agreement with the model of Fan et al. [18].

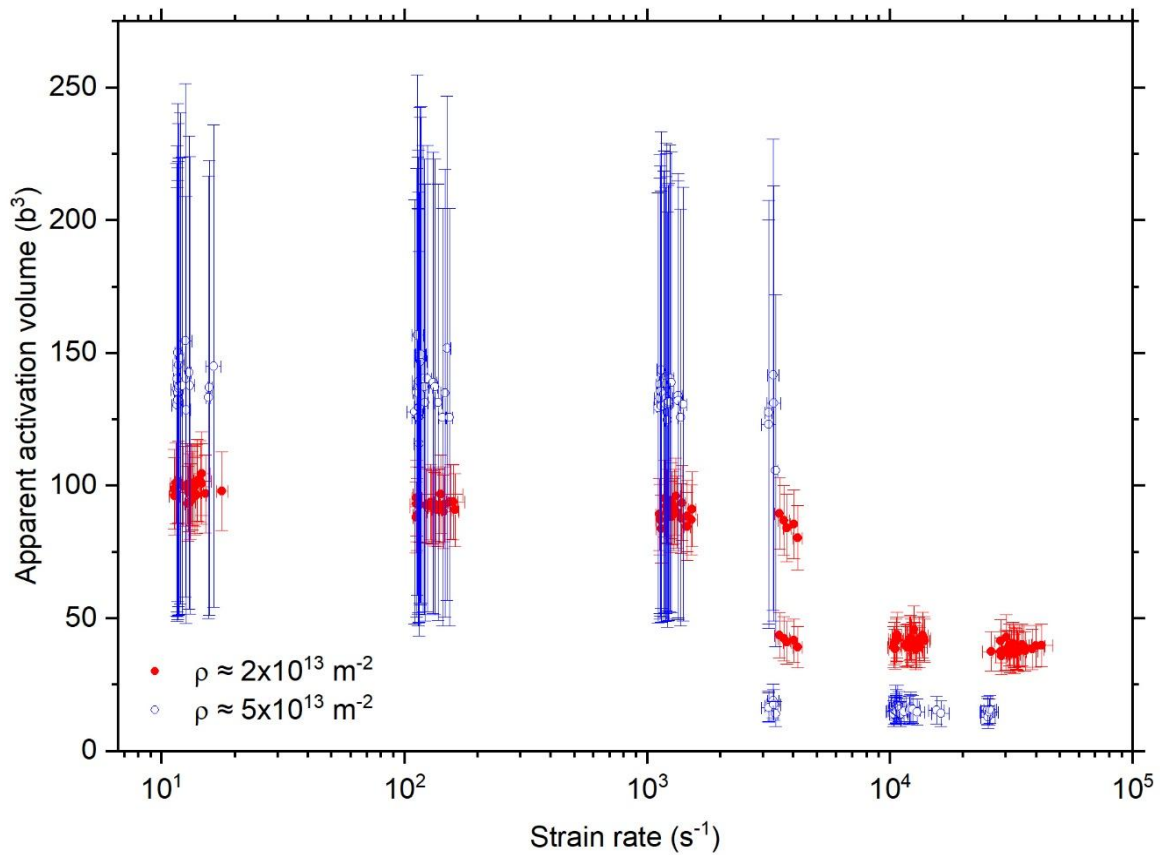

*Figure 6: Activation volume plotted against strain rate for the low dislocation density and high dislocation density sample.*

To investigate the influence of the changed microstructure after deformation on the mechanical properties, reloading experiments were performed in a single-crystal area of the specimen. A first indent at various strain rates ranging from 10 s$^{-1}$ – 25 000 s$^{-1}$ was performed, followed immediately by a second indent at the same location with a strain rate profile of 10 s$^{-1}$. The resulting real strain rate of the reloading indent varies, however, and approaches 5 s$^{-1}$, since the experiment does not start at the initial surface of the material but in an indent. Because of this, and since the final indentation depth is deeper and therefore less prone to the indentation size effect than the indents analyzed in Figure 4 (a), the resulting hardness for the reloading is lower than the 10 s$^{-1}$ indents in Figure 4 (a). For the strain rates between 10 s$^{-1}$ – 1 000 s$^{-1}$, the hardness scatters around the same value. However, for both the high and low dislocation density samples, the hardness starts to increase at ~3 000 s$^{-1}$, matching the upturn regime of Figure 4 (a). If there were no change in microstructure underneath the indent, the hardness for all reloading indents would be the same because all external parameters of the reloading indent are identical. This is a strong indication that the high strain rate deformation leads to changes in microstructure and thus the mechanical properties are changing because of these changes in microstructure. Additionally, the slopes of the hardness increase are within the standard deviation of the slopes observed in the inverse Norton plot in Figure 4 (a). This indicates that a change in microstructure is mainly responsible for the upturn in hardness, and the slope represents a rate dependent microstructure sensitivity instead of the usually measured strain rate sensitivity, consistent with the theory and findings of Follansbee [9], who called it the "rate sensitivity of structure evolution". According to this theory, the ratio of dislocation nucleation to dynamic recovery of dislocations is higher at high strain rates, leading to increased numbers of dislocations after high strain rate deformation [8].

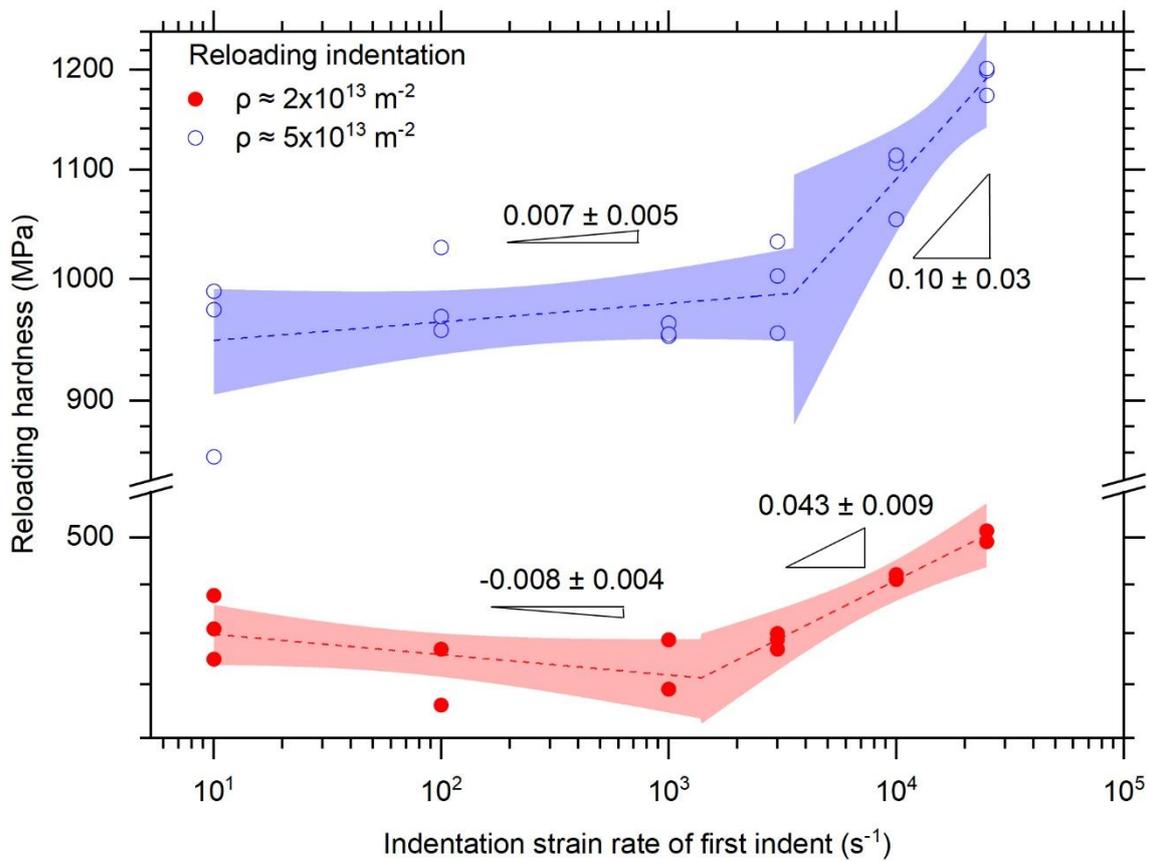

*Figure 7: Hardness of reloading indents at 10 s$^{-1}$ immediately following the first indents of various high indentation strain rates. The dashed line is a piecewise exponential fit, the shaded area corresponds to the 95 % confidence interval.*

We will assume for now that the microstructure change is purely an increase in dislocation density even though in the literature it was shown that the deformed microstructure varied in case of impact testing in single crystal Cu compared to quasi-static indentation. While after quasi-static deformation edge dislocations and LAGBs were found, the high strain rate deformation resulted in partial dislocations and high-angle boundaries, which could influence the hardness as well [52]. So, a convoluted effect or an increased dislocation density as well as grain boundaries that are created by the high strain rate deformation is expected. However, in our case, the assumption is that the dislocation density is the main contributor to the increase in hardness, as the main contributor in the bulk of the deformation is the dislocation network. Using Taylor's hardening equation, the increase in dislocation density can be calculated from the reloading hardness using [45]

$$\tau \cong \tau_0 + G\alpha b\sqrt{\rho}, \qquad (2)$$

where $\tau$ is the shear stress, $\tau_0$ is the intrinsic lattice resistance to dislocation motion, $G$ is the shear modulus, $\alpha$ is a material dependent constant, $b$ is the Burgers vector and $\rho$ is the dislocation density. Since for fcc metals the intrinsic lattice resistance is negligible [52] and since in single crystals the hardness is related to the shear stress by a constraint factor, the hardness change can be linearly related to the dislocation density by:

$$\frac{H_{\text{high}}}{H_{\text{low}}} \propto \sqrt{\frac{\rho_{\text{high}}}{\rho_{\text{low}}}}, \qquad (3)$$

where $H_{low}$ and $\rho_{low}$ are the hardness and dislocation density before the upturn and $H_{high}$ and $\rho_{high}$ are the respective values after the upturn. To obtain the increase in dislocation density, the ratio of the hardness after and before the upturn was calculated. For $H_{\text{low}}$, all hardness data before the upturn were averaged. For $\rho_{low}$ the GND density from the EBSD measurements was used. This leads to an increase of $\rho$ in the low dislocation density sample of (13 ± 6) % at 10 000 s$^{-1}$ and to a (22 ± 7) % increase at 25000 s$^{-1}$. The increase of $\rho$ in the high dislocation density sample is (27 ± 14) % and (51 ± 15) % at 10 000 s$^{-1}$ and 25 000 s$^{-1}$, respectively. Unfortunately, these increases in dislocation density are not within the precision of EBSD measurements conducted within the small size of the indent. The findings strongly indicate that the microstructure underneath the indents at strain rates above the critical upturn differs significantly from nanoindents at lower strain rates, supporting earlier theories by Follansbee [9]. An extensive study of multiple lift outs from the indents using transmission electron microscopy could help in the future to understand the deformation mechanisms further. For example, annular bright field scanning transmission electron microscopy images can be used to estimate the dislocation density underneath the indents, similar to the work in [17].

## Conclusion

In this work, we performed high strain rate nanoindentation (10 s$^{-1}$ – 25 000 s$^{-1}$) on two copper samples with different dislocation densities of ~5 x 10$^{13}$ m$^{-2}$ and ~2 x 10$^{13}$ m$^{-2}$, targeting regions near and far from a LAGB with tilt and twist components. We find that the LAGB does not significantly affect hardness or the strain rate at which the upturn in strength occurs within the experimental scatter. However, the slip lines were found to stop on the LAGB, indicating that the LAGB is a barrier to dislocation motion, a result confirmed by molecular dynamics simulations. In contrast, hardness increases with dislocation density in agreement with Taylor hardening in the bulk, and an upturn in rate-dependent hardness emerges above ~3 000 s$^{-1}$, consistent with predictions from dislocation density-based DDD and MD simulations reported in literature. Changes in the activation volume at the upturn and reloading experiments further suggest that local microstructural changes beneath the indent contribute to this upturn, supporting earlier rate-dependent microstructure sensitivity theoretical hypotheses by Follansbee [9].

## Acknowledgements & Funding

D.S. was supported by the Alexander von Humboldt Foundation. R.R. would like to acknowledge funding from the ERC (Grant agreement No. 101078619; AMMicro) and Eurostars Project HINT (01QE2146C). Views and opinions expressed are, however, those of the author(s) only and do not necessarily reflect those of the European Union or the European Research Council. Neither the European Union nor the granting authority can be held responsible for them. G. D. and L. K. B. acknowledge the support by the Deutsche Forschungsgemeinschaft (DFG) within project B06 of Collaborative Research Center (SFB) 1394 "Structural and Chemical Atomic Complexity - from defect phase diagrams to material properties", project number 409476157.

## Data availability

The experimental data is available on Zenodo and can be downloaded from: https://doi.org/10.5281/zenodo.18831995

## Conflict of interests

The authors declare no competing interests.

## Author contributions

Hendrik Holz: Conceptualization, Methodology (Sample preparation, EBSD measurement, nanoindentation), Formal analysis (EBSD data, nanoindentation data), Investigation, Writing - Original Draft, Writing - Review & Editing, Visualization; Lalith Kumar Bhaskar: Conceptualization, Methodology (nanoindentation), Formal analysis (nanoindentation data), Investigation, Writing - Review & Editing; Tobias Brink: Methodology (Molecular statics and dynamics simulation), Formal analysis (Molecular statics and dynamics simulation), Investigation, Writing - Review & Editing; Dipali Sonowane: Conceptualization, Writing - Review & Editing; Gerhard Dehm: Conceptualization, Writing - Review & Editing, Supervision; James Best: Conceptualization, Writing - Review & Editing, Supervision; Rajaprakash Ramachandramoorthy: Conceptualization, Writing - Review & Editing, Supervision

# Declaration of generative AI and AI-assisted technologies in the writing process

During the preparation of this work the authors used ChatGPT 5 in order to improve the readability and language of the manuscript. After using this tool/service, the authors reviewed and edited the content as needed and take full responsibility for the content of the published article.

# Supplementary information

Table S1: Lateral forces measured at different strain rates on all samples. The high dislocation density sample is shown twice, as there was a second measurement session on the Grain [6 4 11] using a different piezo tube actuator, resulting in different lateral forces.

| Sample | Strain rate ($s^{-1}$) | Max lateral force at peak load (mN) | Max relative lateral force at peak load (%) | Average lateral force at peak force (mN) | Average relative lateral force at peak load (%) |
|---|---|---|---|---|---|
| High dislocation density sample | ~10 | 2.1 | 6.5 | 1.5 | 5.1 |
| | ~100 | 2.3 | 5.4 | 1.5 | 4.6 |
| | ~1 000 | 1.4 | 4.4 | 1.2 | 3.7 |
| | ~10 000 | 0.7 | 1.8 | 0.4 | 1.2 |
| High dislocation density sample (Separate measurement session for Grain [6 4 11] using different piezo tube actuator) | ~10 | 0.4 | 1.1 | 0.0 | 0.0 |
| | ~100 | 0.1 | 0.2 | 0.1 | 0.2 |
| | ~1 000 | 0.2 | 0.6 | 0.2 | 0.5 |
| | ~3 000 | 0.4 | 0.9 | 0.2 | 0.5 |
| | ~10 000 | 0.5 | 1.0 | 0.0 | 0.0 |
| | ~25 000 | 0.3 | 0.5 | 0.0 | -0.1 |
| Low dislocation density sample | ~10 | 1.1 | 5.3 | 0.7 | 4.3 |
| | ~100 | 1.1 | 5.3 | 0.8 | 4.4 |
| | ~1 000 | 1.1 | 6.0 | 0.9 | 5.2 |
| | ~3 000 | 0.9 | 4.6 | 0.8 | 4.4 |
| | ~10 000 | 0.6 | 2.9 | 0.5 | 2.5 |
| | ~25 000 | 0.4 | 1.8 | 0.3 | 1.2 |

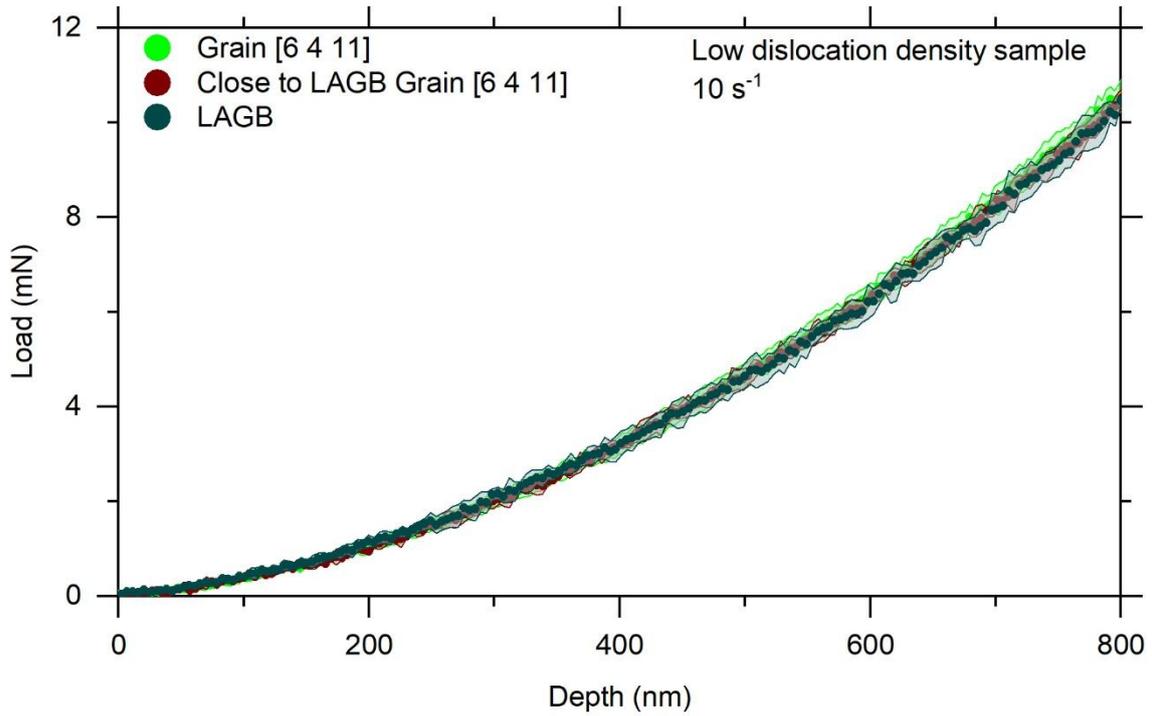

*Figure S1: Load-displacement data of nanoindentation experiments after postprocessing corrections from the low dislocation density sample showcasing the small standard deviation between nanoindentation results in the single grain, close to the LAGB and on the LAGB. Exemplary curves of 10 s$^{-1}$ data. The circles represent the average of at least 3 experiments.*

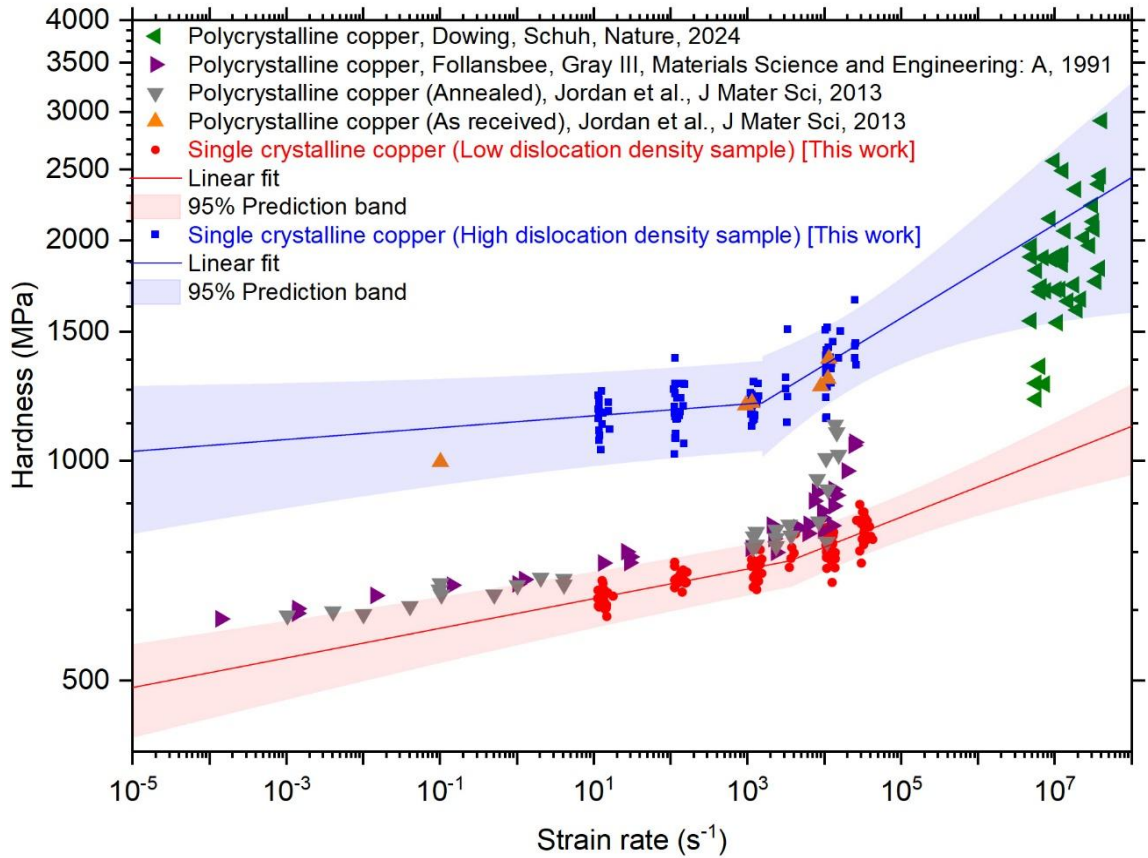

Figure S2: Measured hardness over strain rate compared with literature data. The uniaxial yield stresses from Follansbee and Gray [2] and Joradan et al. [3] were multiplied by a constraint factor of 3 to obtain hardness. The fit for the prediction bands for each data set of this work is performed by a piecewise fit as described in supplemental note 2.

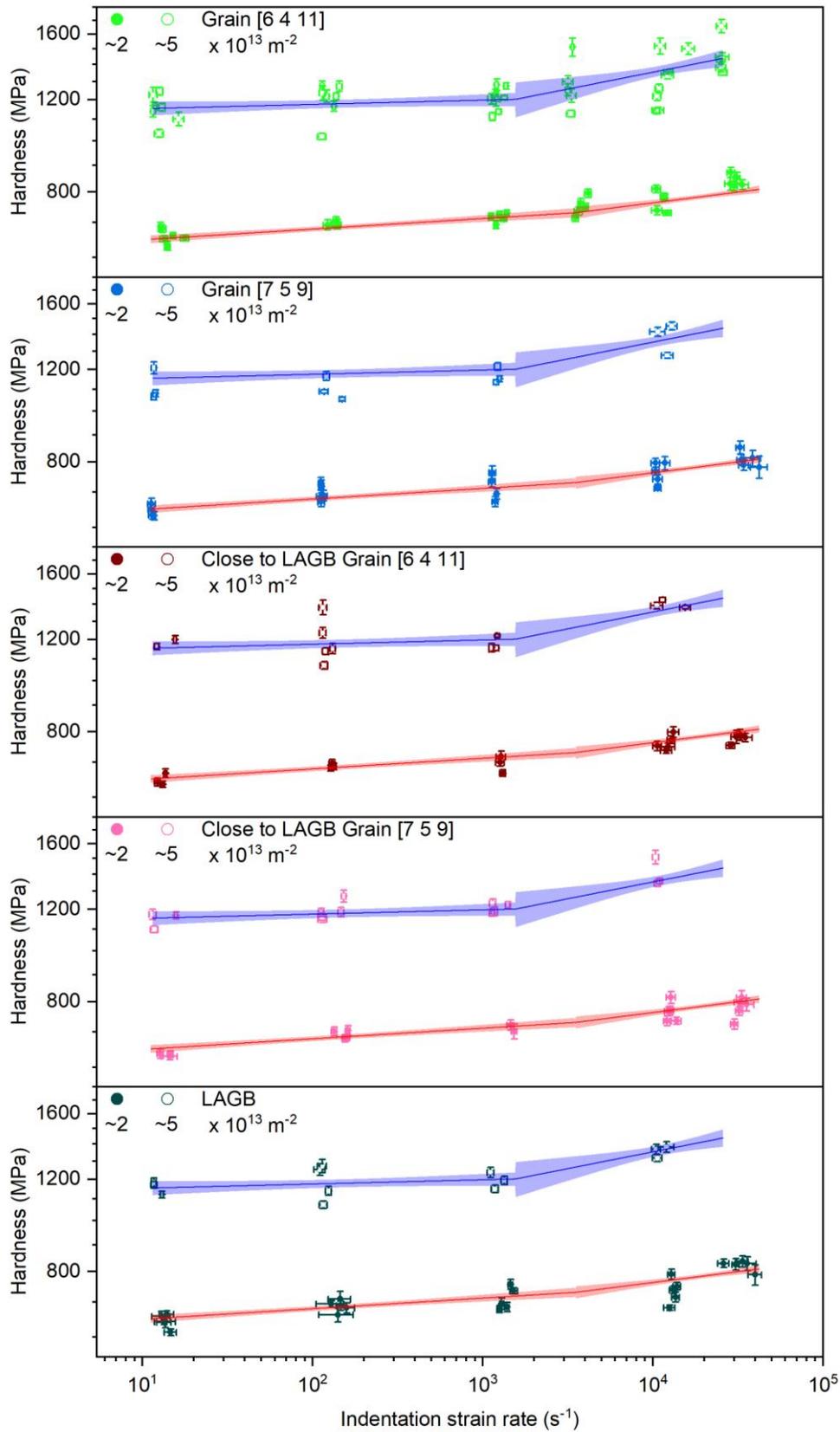

Figure S3: Rate sensitivity of copper with varying dislocation density and position. The fit shown in all figures is the result of the piecewise fit including all datapoints and is in these figures as a guide for the eye.

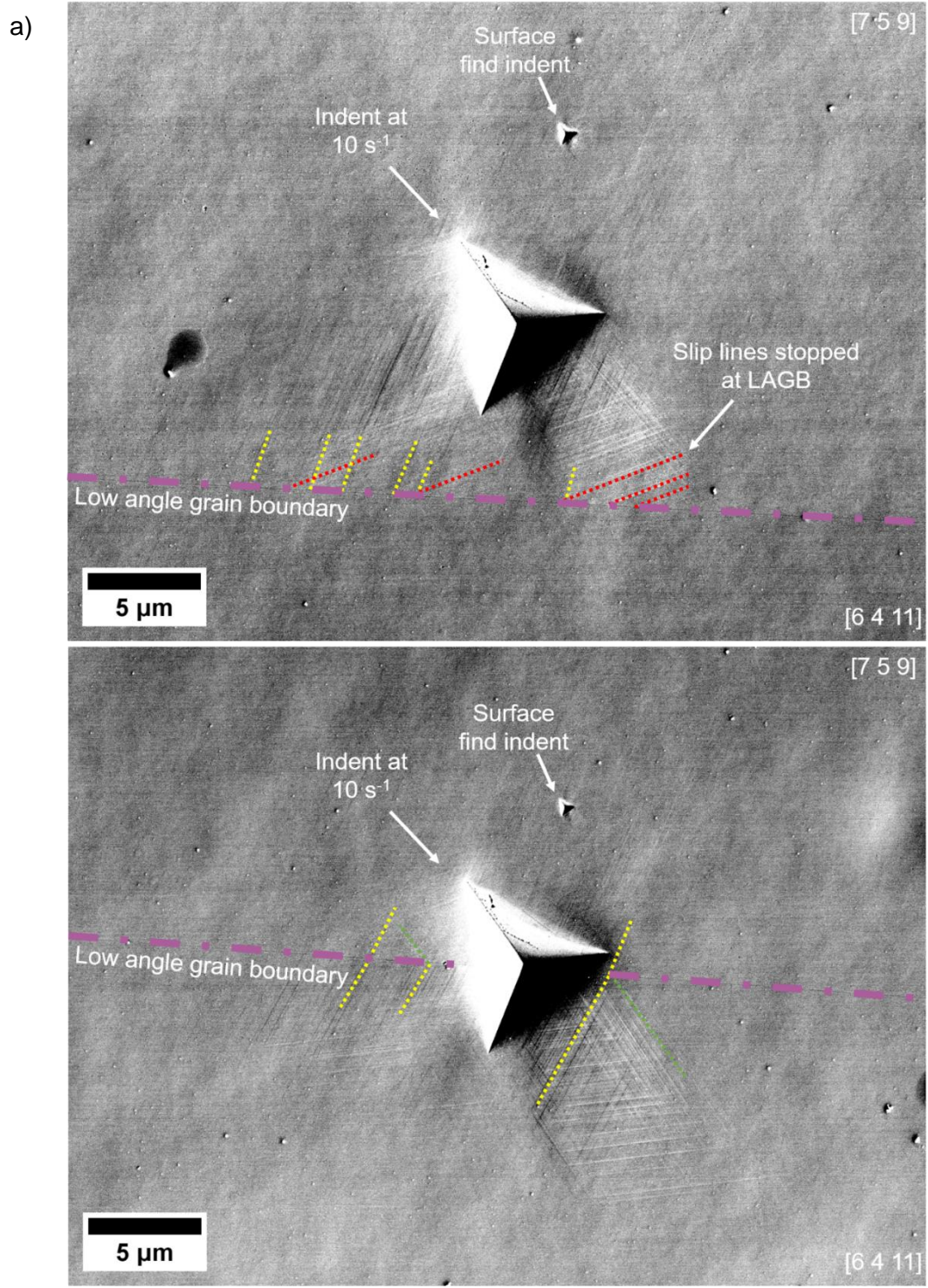

*Figure S4: SEM micrograph of indents performed at 10 s$^{-1}$ in the low dislocation density sample. (a) Indent located close to the low angle grain boundary showing slip lines stopping at the boundary. (b) Indent located at the low angle grain boundary showing different slip line orientation in the respective single grains.*

*The contrast and brightness were enhanced to make the slip lines and the LAGB more visible. Dashed lines showing the LAGB and slip lines were added as guide for the eye.*

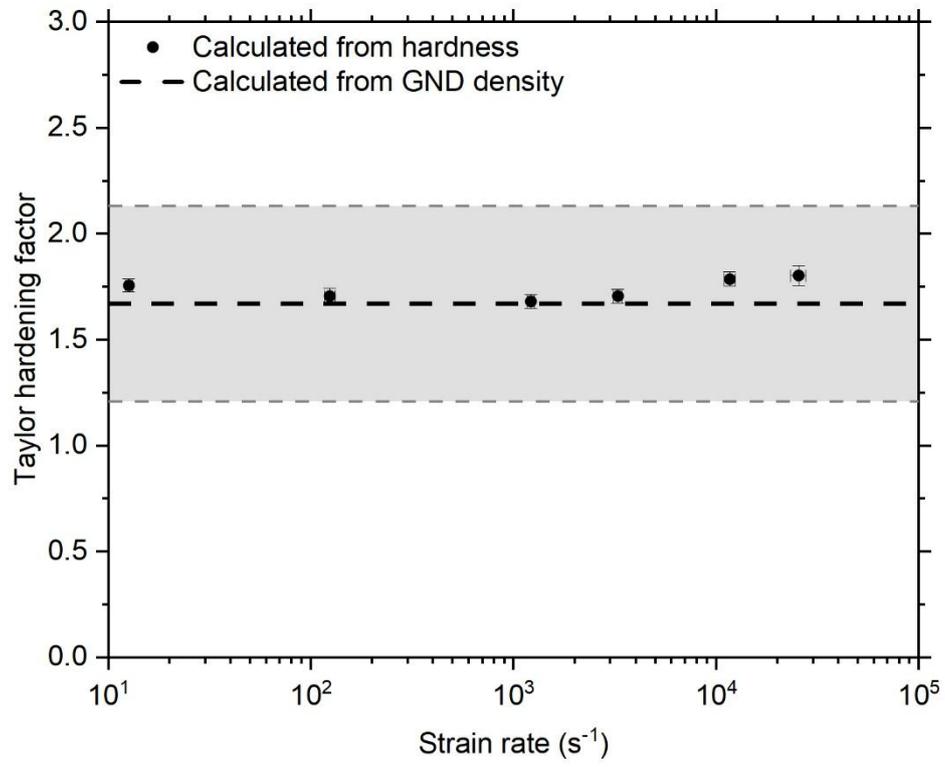

*Figure S5: Taylor hardening rate calculated from dislocation density (dashed line) and difference in hardness (dots). The grey background represents the standard deviation calculated using the GND density data.*

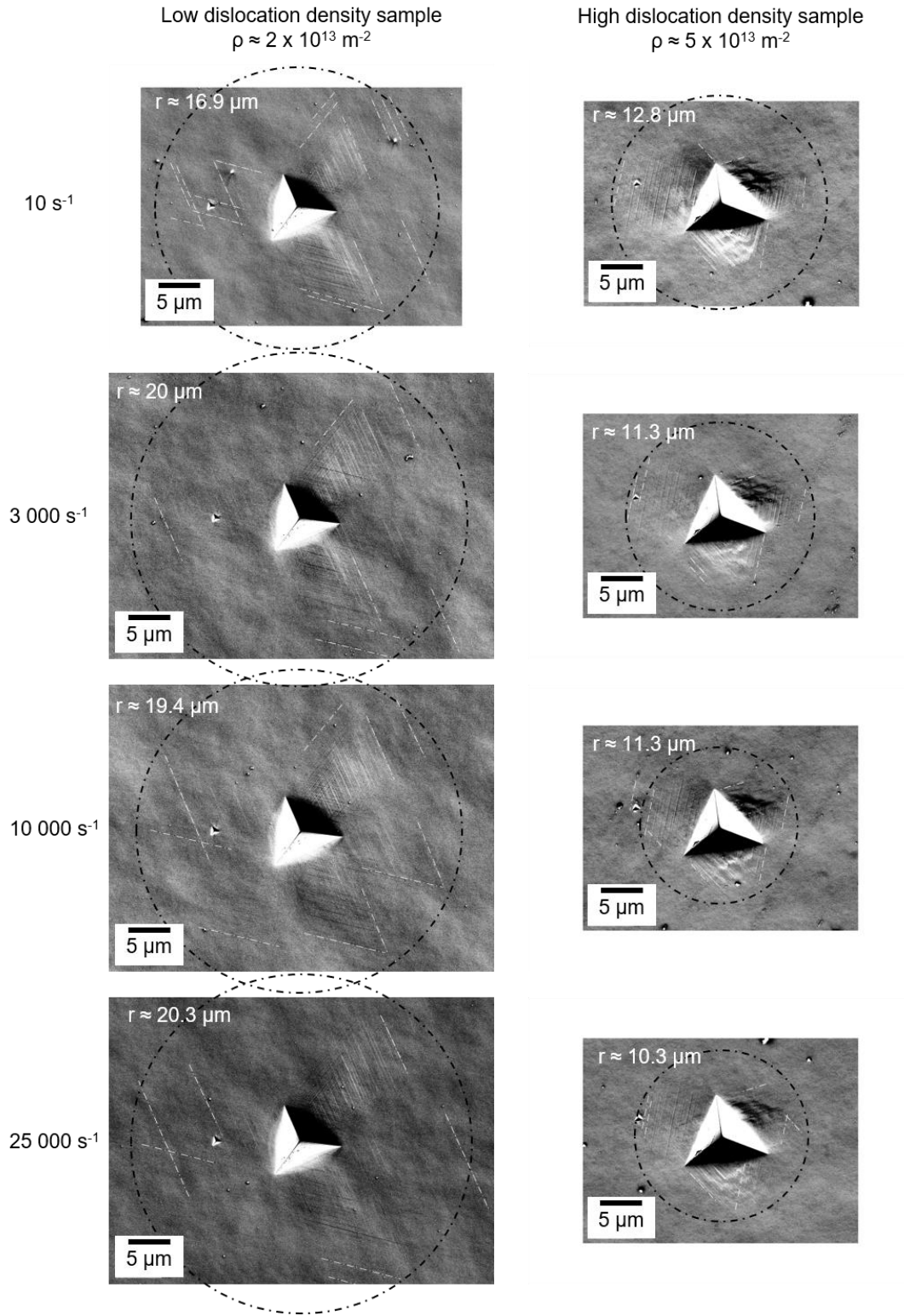

*Figure S6: Micrographs of indents from different strain rates in the [6 4 11] oriented single crystal showing a clear difference in the travel distance of the visible slip lines from the nanoindents taken on samples with low and high dislocation density. The radius of the circle was chosen so that a majority of slip lines stay within the circle.*

*The contrast and brightness were enhanced to make the slip lines more visible. Dashed lines showing slip lines were added as a guide for the eye.*

Supplemental Note 1:

In Taylor hardening [4], we assume that dislocations have to pass a forest of perpendicular dislocations with average distance $R$, which represent obstacles. This leads to an Orowan stress due to bowout. We can make a similar derivation for LAGBs, by assuming that the LAGB dislocations are arranged in a grid with distance $R$ (Figure S6). This assumption holds true for our mixed twist/tilt LAGB (see Figure 2 in the main text), while a pure tilt LAGB would have an array of parallel dislocations, for example.

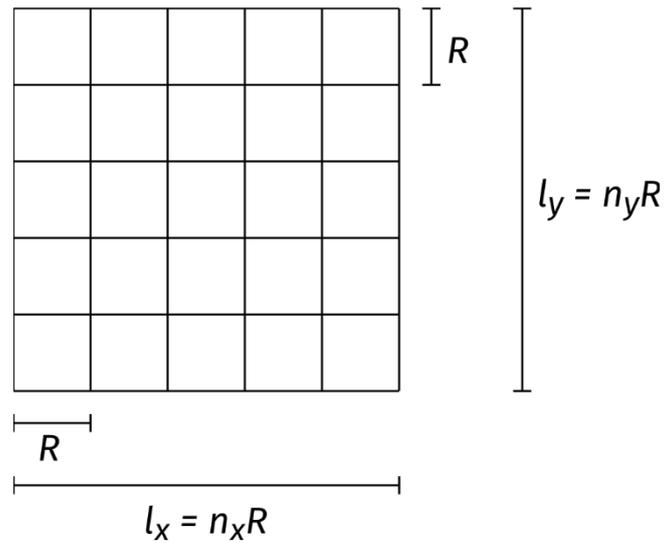

*Figure S7: Schematic representation of the grid representing a simplified version of dislocations in a mixed tilt/twist LAGB.*

For an assumed rectangular grain boundary area with side lengths $l_x$ and $l_y$, we obtain an area $A = l_x l_y$. Given that we have $n_x$ vertical dislocations and $n_y$ horizontal ones, we can also write $A = n_x n_y R^2$. The total dislocation length is consequently $L = n_x l_y + n_y l_x = 2 n_x n_y R$. This leads to an area density $\rho_{\text{area}} = \frac{L}{A} = \frac{2}{R}$. This value is also valid for non-rectangular grain boundary regions.

With an Orowan stress of $\tau = \frac{Gb}{R}$, we find a Taylor-like expression of $\tau = \frac{1}{2} Gb \rho_{\text{area}}$. With the values from the simulation (see Figure 2 in the main text), this model yields $R \cong 0.8$ nm, $\rho_{\text{area}} \cong 2.57 \times 10^9 /\text{m}$, and thus $\tau \cong 6.3$ GPa. This value is expectedly quite high, because of the dense dislocation spacing in our particular LAGB.

Supplemental Note 2:

The strain-rate dependence of the flow stress was described by a two-segment piecewise power-law function:

$$\sigma(\dot{\varepsilon}) = \begin{cases} A\dot{\varepsilon}^{m_1}, & \dot{\varepsilon} < \dot{\varepsilon}_c \\ A\dot{\varepsilon}_c^{m_1-m_2}\dot{\varepsilon}^{m_2}, & \dot{\varepsilon} \geq \dot{\varepsilon}_c \end{cases}$$

where A is a pre-factor, $m_1$ and $m_2$ are the strain-rate sensitivity exponents below and above the critical strain rate $\dot{\varepsilon}_c$, respectively, and $\dot{\varepsilon}_c$ is the fitted breakpoint strain rate at which the slope changes.

To differentiate the errors in prediction from the fits, the RMSD is used. It is calculated using $RMSD = \sqrt{\frac{1}{N}RSS}$ where N is the number of samples, and RSS is the residual sum of squares.

The values for the piecewise fit compared to a single fit are shown in Table S2:

Table S2: Result of the root square mean difference analysis enforcing a single fit and using a two piecewise fit.

| Sample | Fit | RSS [MPa$^2$] | N | RMSD [MPa] |
|---|---|---|---|---|
| High dislocation density | Single fit | 8.99438 × 10$^5$ | 93 | 98.3 |
| | Piecewise fit | 6.14773 × 10$^5$ | | 81.3 |
| Low dislocation density | Single fit | 9.82627 × 10$^5$ | 117 | 29.0 |
| | Piecewise fit | 9.03572 × 10$^5$ | | 27.8 |

This showcases that the piecewise fit is superior to the single fit in both instances.